\documentclass[a4paper,10pt]{article}
\usepackage[utf8]{inputenc}
\usepackage{color}
\usepackage{cite}
\usepackage{hyperref} 
 \usepackage{latexsym}
  \usepackage{amssymb}
  \usepackage{graphicx}
   \DeclareGraphicsExtensions{.pdf .jpg .png .gif.pdf}
\usepackage[T1]{fontenc}
 \usepackage[english]{babel}
  \usepackage{amstext}
  \usepackage{amsmath}
  \usepackage{times} 
  \usepackage{amsfonts}
 \date{}

\title{ Nonlinear dynamics of system oscillations modeled by a 
forced Van der Pol generalized oscillator }
\author{ L. A. Hinvi\footnote{laurent.hinvi@imsp-uac.org} ,C. H. Miwadinou\footnote{clement.miwadinou@imsp-uac.org, hodevewan@yahoo.fr},  
A. V. Monwanou\footnote{movins2008@yahoo.fr} and \\
J. B. Chabi Orou\footnote{Author to whom correspondence should be addressed: jchabi@yahoo.fr}}

\begin{document}

\maketitle

\begin{abstract}
This paper considers  the   oscillations modeled by 
a forced  Van der Pol generalized oscillator.
These  oscillations are described by a nonlinear differential equation of the form
$ \ddot{x}+x-\varepsilon\left(1-ax^2-b\dot{x}^2\right)\dot{x}=E\sin{{\Omega}t}.$ 
The amplitudes of the forced harmonic,  primary resonance
superharmonic and subharmonic oscillatory states are obtained using the harmonic balance 
technique and the multiple time scales methods. We obtain also the hysteresis and jump phenomena in the system oscillations.
 Bifurcation sequences displayed by the model for each
type of oscillatory states are performed numerically through the fourth-order Runge- Kutta scheme.
\end{abstract}
 
\section{Introduction}
Many problems in physics,  chemistry,  biology,  etc.,  are related to nonlinear
self-excited oscillators (\cite{0} - \cite{1}). 
 Thus, Balthazar Van der Pol $(1889-1959)$ was a Dutch electrical engineer who
initiated modern experimental dynamics in the laboratory during the $1920's$ and
$1930's$. He, first, introduced his (now famous) equation in order to describe
triode oscillations in electrical circuits, in $1927$ (\cite{2}-\cite{12}). 
The mathematical model for the
system is a well known second order ordinary differential equation with cubic
nonlinearity Van der Pol equation. Since then thousands of papers have
been published achieving better approximations to the solutions occurring in
such non linear systems. The Van der Pol oscillator is a classical example of
self-oscillatory system and is now considered as very useful 
mathematical model that can be used in much more complicated and modified systems.
But, why this equation is so important to mathematicians, physicists and
engineers and is still being extensively studied?

During the first half of the twentieth century, Balthazar Van der
Pol pioneered the fields of radio and telecommunications \cite{2}.
In an era when these areas were much less advanced than they are
today, vacuum tubes were used to control the flow of
electricity in the circuitry of transmitters and receivers. Contemporary with Lorenz, Thompson,
and Appleton, Van der Pol, in $1927$, experimented with oscillations in a vacuum tube triode circuit and
concluded that all initial conditions converged to the same periodic orbit of finite
amplitude. Since this behavior is different from the behavior of solutions of linear
equations, Van der Pol proposed a nonlinear differential equation
\begin{eqnarray}
\ddot{x} + x-\varepsilon(1-x^{2})\dot{x}=0, \label{eq0}
\end{eqnarray}
commonly referred to as the (unforced) Van der Pol equation \cite{3}, as a model for
the behavior observed in the experiment. In studying the case $\varepsilon \gg1$, Van der Pol
discovered the importance of what has become known as relaxation oscillations 
 (\cite{4}, \cite{10}).  Van der Pol went on to propose a version of (\ref{eq0}) that includes a periodic
forcing term \cite{8}:
\begin{eqnarray}
\ddot{x} + x-\varepsilon(1-x^{2})\dot{x}=E\sin{\Omega}t. \label{eq00}
\end{eqnarray}
Many systems  have characteristics of two types of oscillators   and whose equation presents a
combination of terms of these oscillators. Thus, we  have the systems characterized by Van der Pol and 
Rayleigh oscillators  namely the  Van der Pol generalized oscillator \cite{011} or hybrid Van der pol-Rayleigh oscillator  modeled by 
 \begin{eqnarray}
 \ddot{x}+x-\varepsilon\left(1-ax^2-b\dot{x}^2\right)\dot{x}=0. \label{eq000}
\end{eqnarray}
 The Van der Pol  generalized oscillator  as all oscillator, models many physical systems. 
 Thus, we have  \cite{12} who 
 models a bipedal robot locomotion   with this oscillator '' known as Hybrid 
Van der Pol-Rayleigh oscillators''.
 
In the present paper, We considered  the forced Van der Pol generalized oscillator equation \cite{011} 
 \begin{eqnarray}
 \ddot{x}+x-\varepsilon\left(1-ax^2-b\dot{x}^2\right)\dot{x}=E\sin{\Omega t}, \label{eq1}
\end{eqnarray}
where $a$, $b$  are  positifs
cubic nonlinearities,  $\epsilon \ll 1$ is  damping parameters while   $E$  and  $\Omega$ stand for the amplitude and  the pulsation  of 
the external excitation.

We concentrate our studies on the  equation of motion (\ref{eq1}),  the
resonant states,  the chaotic behavior. Through these studies,  we found the effects of differents
parameters in general.

The paper is structured as follows: Section $2$  gives an analytical treatment of equation (\ref{eq1}).
Amplitude of the forced harmonic oscillatory 
states is obtained with harmonic-balance method \cite{13}. Section $3$ investigate using multiple time-scales
method \cite{14} the resonant cases and the stability conditions are found by the perturbation method 
\cite{13}. The section $4$ evaluates  bifurcation and chaotic behavior by numerically
simulations of equation (\ref{eq1}). Section $5$ deals with conclusions.

\section{Amplitude of the forced harmonic oscillatory states}
Assuming that the fundamental component of the solution and the external 
excitation have the same period,  the amplitude of harmonic oscillations can be
tackled using the harmonic balance method ( \cite{13}). For this purpose,  we express
its solutions as
\begin{eqnarray}
 x &=& A \sin{\Omega t}  +\xi \label{eq2}
\end{eqnarray}
where $A$ represents the amplitude of the oscillations and $\xi \ll1$
a constant. Inserting this solution (\ref{eq2})in (\ref{eq1}) we obtain
\begin{eqnarray}
 (1-\Omega^2)A\sin{\Omega t} +\xi+\varepsilon\left[(1-a{\xi}^2)A \Omega -\frac{a A^3\Omega}{4} 
 -\frac{3 bA^3{\Omega}^3}{4}\right] \cos{\Omega t}+\nonumber\\
 +\varepsilon \frac{\left(a A^3\Omega- bA^3{\Omega}^3\right)}{4} \cos{3\Omega t}-\varepsilon \xi a A^2 \sin{2\Omega t}=
 E\sin{\Omega t}.\label{eq3}
\end{eqnarray}
 By equating the constants and the coefficients of
 $\sin{\omega}$ and $\cos{\omega}$,  we have
 \begin{eqnarray}
  (1-\Omega^2)A&=&E, \label{eq4}\\
  \xi&=&0,\label{eq5}\\
  \varepsilon\left[(1-a{\xi}^2)A \Omega -\frac{a A^3\Omega}{4} -\frac{3 bA^3{\Omega}^3}{4}\right]&=&0\label{eq6}
 \end{eqnarray}
$\Longleftrightarrow$
 \begin{eqnarray}
  (1-\Omega^2)^2A^2&=&E^2,\label{eq9}\\ 
  \varepsilon^2\left[(1-a{\xi}^2)A \Omega -\frac{a A^3\Omega}{4} -\frac{3 bA^3{\Omega}^3}{4}\right]^2&=&0. \label{eq11} 
 \end{eqnarray}
 We combined  the Equations (\ref{eq9} -\ref{eq11}) and we obtain
 \begin{equation}
 \left[(1-\Omega^2)^2 -\varepsilon^2\Omega^2\right] A^2 +\frac{\varepsilon^2\Omega^2}{2}(a+3b\Omega^2)A^4 
 -\frac{\varepsilon^2\Omega^2}{16}(a+3b\Omega^2)^2A^6=E^2.\label{eq15}                  
\end{equation}
\begin{figure}[htbp]
\begin{center}
 \includegraphics[width=12cm,  height=5cm]{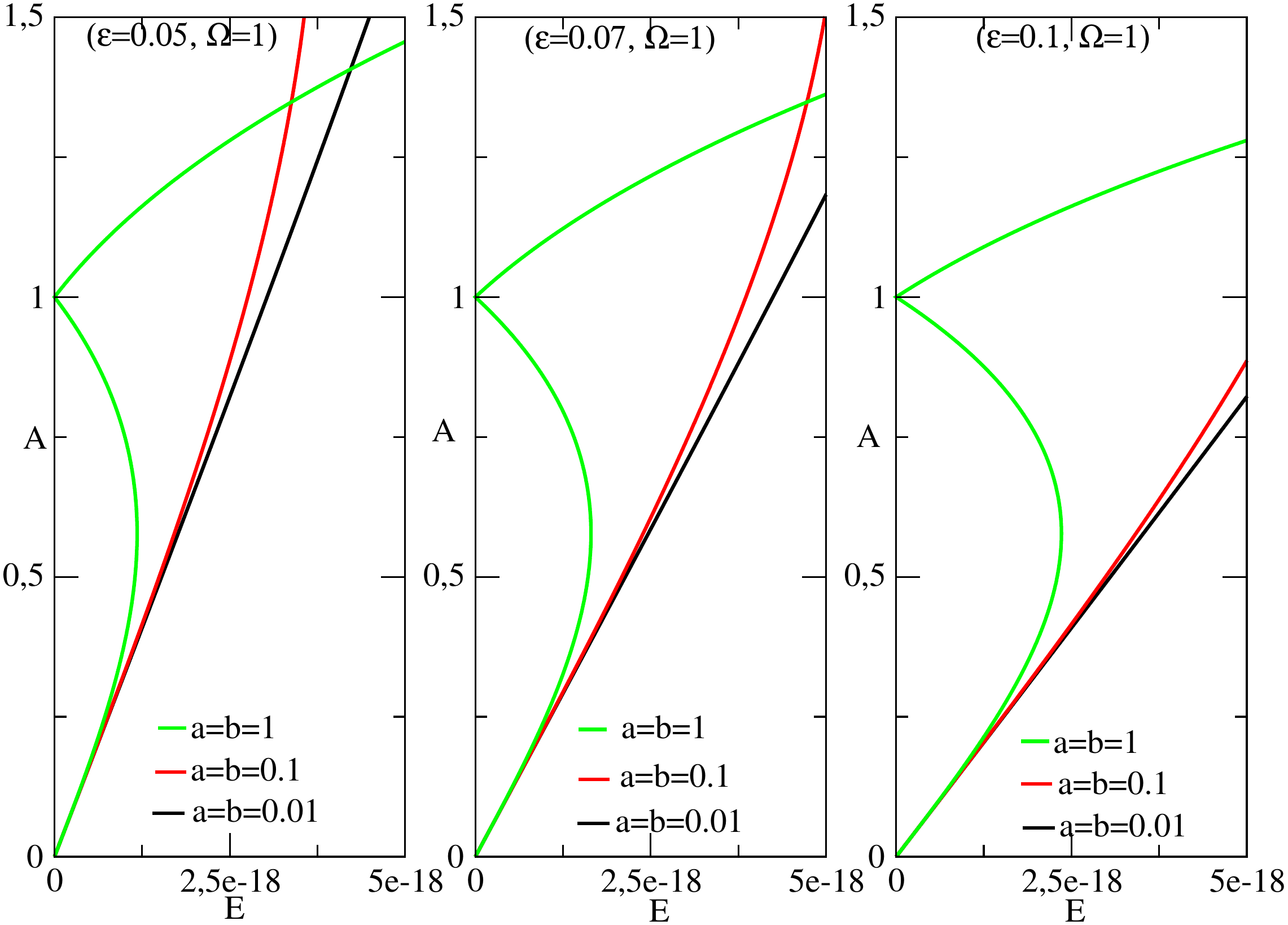}
\end{center}
\caption{Effects of  the parameters $a, b, \varepsilon$ on the amplitude-response curves with 
$ \Omega=1$.}
\label{fig:1}
\end{figure}
\begin{figure}[htbp]
\begin{center}
 \includegraphics[width=12cm,  height=6cm]{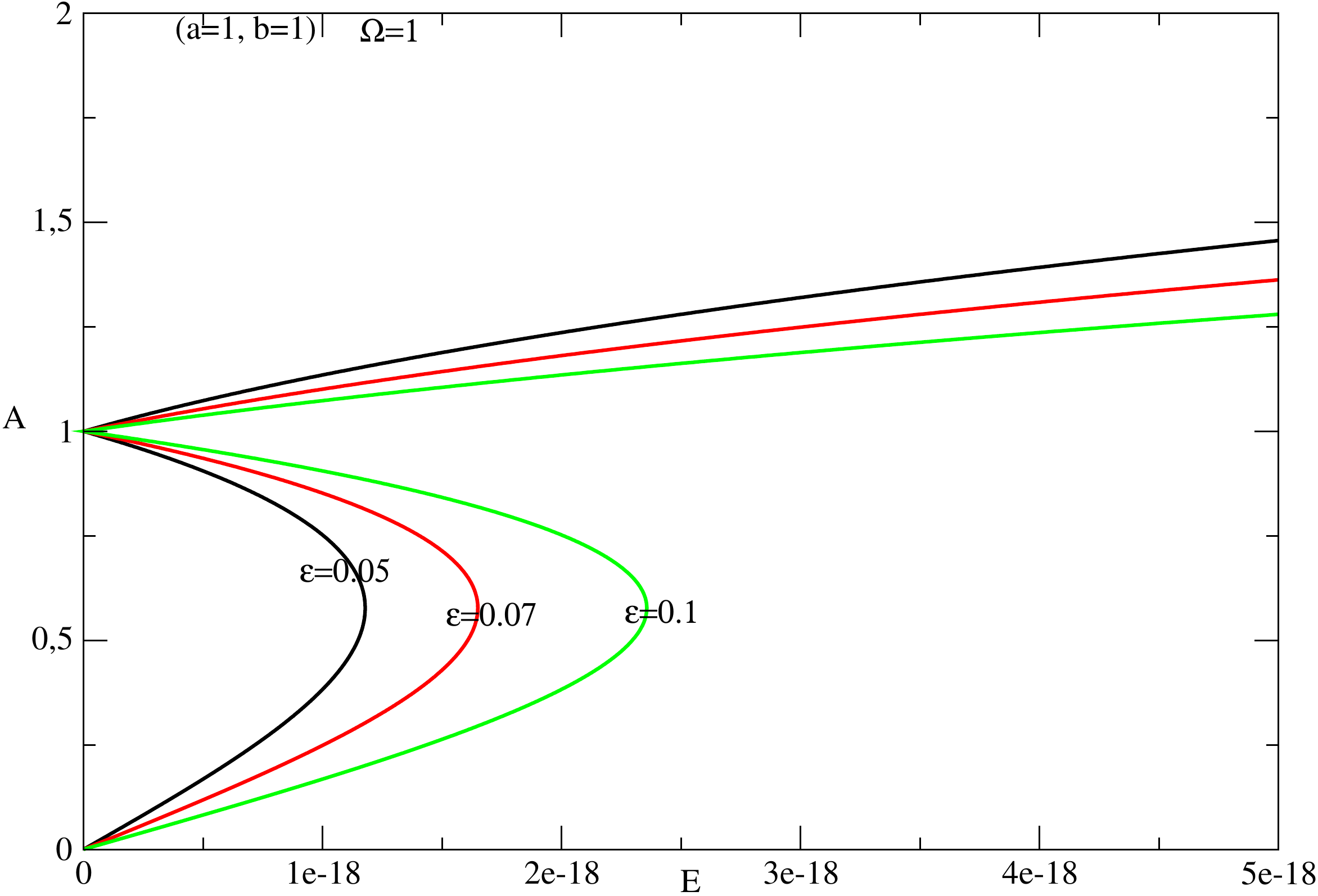}
\end{center}
\caption{Effects of  the parameter $\varepsilon$   on the amplitude-response curves with 
$a=b=1, \Omega=1 $.}
\label{fig:2}
\end{figure}
\newpage
 The figures \ref{fig:1} and \ref{fig:2} show  the effects of  the parameters $a, b, \varepsilon$   
on the amplitude-response curves $A(E)$. Through these figures, we note that for small  value of Rayleigh coefficient $a$ and Van der Pol coefficient
 $b$ the amplitude-response is linear and when its increase, the hystereris and jump phenomenas appear. The gape of  jump phenomenon   
descreases when  the damping parameter $\varepsilon$ increases but hysteresis and jump phenomena persist.

\section{Resonant states}
We investigate the differents resonances with the Multiple time scales Method $MSM$.
In such a situation,  an approximate solution is generally sought as follows:
\begin{eqnarray}
 x(\epsilon,  t)&=&x_{0}(T_{0},  T_{1} )+ \epsilon x_{1}(T_{0},  T_{1} )+ \dots \label{eq18}
\end{eqnarray}
With $T_{n}={\epsilon}^{n}t$.

The derivatives operators can now be rewritten as follows:
 \begin{eqnarray}
\left\{
 \begin{array}{cl}
\frac{d}{dt}=D_{0} +\epsilon D_{1}+\dots \label{eq19}             
\\
\\
\frac{d^{2}}{{dt}^{2}}=D_{0}^{2}+2\epsilon D_{0}D_{1} +\dots  
\end{array}
\right
.\end{eqnarray}
where ${D^{m}_{n}}=\frac{\partial^{m}}{\partial{T_{n}^{m}}}$.

\subsection{Primary resonant state}
In this state,  we put that 
$E\propto \epsilon \tilde{E}$.
 The closeness between both internal and external frequencies is given by 
 $\Omega = 1+\epsilon \sigma$. Where $\sigma=o(1)$ is the detuning parameter, the internal frequencie is $1$. Inserting (\ref{eq18}) 
 and (\ref{eq19})
 into (\ref{eq1}) we obtain:
 \begin{eqnarray}
   \left(D_{0}^{2}+2\epsilon D_{0}D_{1}\right)(x_{0}+ \epsilon x_{1}) +
   x_{0}+ \epsilon x_{1}-{\epsilon} E\sin{\Omega t}
  = \nonumber\\  \epsilon \left[ 1-a{\left(x_{0}+ \epsilon x_{1}\right)}^{2}-b \left[\left(D_{0} 
  +\epsilon D_{1}\right)(x_{0}+\epsilon x_{1} )\right]^2  \right]
   \left(D_{0} +\epsilon D_{1}\right)(x_{0}+\epsilon x_{1} ). \label{eq21}
 \end{eqnarray}

 Equating  the coefficients of like powers of $\epsilon$ after some algebraic 
 manipulations,  we obtain:
 \begin{eqnarray}
\left\{
 \begin{array}{cl}
 D^{2}_{0}x_{0}+x_{0}=0,\label{eq22}              
\\
\\
D^{2}_{0}x_{1}+x_{1}=E\sin{\Omega}t-2D_{1}D_{0}x_ {0}+ \left( 1-ax_{0}^{2}-b (D_{0}x_{0})^2\right)D_{0}x_{0}.
\end{array}
\right
.\end{eqnarray}
The general solution of  the first equation of system (\ref{eq22}) is 
\begin{eqnarray}
 x_{0}&=&A(T_{1})\exp(jT_{0})+CC,  \label{eq24}
\end{eqnarray}
where $CC$ represents the complex conjugate of the previous terms.
$A(T_{1} )$ is a complex function to be determined from solvability or secular conditions 
of the second equation of system (\ref{eq22}). Thus,  substituting the solution $x_{0}$ in (\ref{eq22})
leads us to the following secular criterion
\begin{eqnarray}
 D^{2}_{0}x_{1}+x_{1}&=&j\left[-2{A'}+ \left(1-a{|A|}^{2}-3b{|A|^{2}}\right)A\right]e^{j T_{0}}+
 j\left(b-a\right)A^3e^{3jT_{0}}+\cr
 &&-\frac{jE}{2}e^{j(1+\epsilon \sigma)t} +CC \label{eq25} \nonumber\\
 &=&j\left[-2{A'}+ \left(1-(a+3b){|A|^{2}}\right)A-\frac{E}{2}e^{j\sigma \epsilon t}\right]e^{j T_{0} }+\cr
 &&+j\left(b-a\right)A^3e^{3jT_{0}}+CC.
\end{eqnarray}
The secular criterion follows
\begin{equation}
 \left[-2{A'}+ A-(a+3b){|A|^{2}}A-\frac{E}{2}e^{j\sigma T_{1}}\right]=0.\label{eq28}
\end{equation}
In polar coordinates,  the solution of  (\ref{eq28}) is
\begin{eqnarray}
 A&=&\frac{1}{2}p(T_{1})\exp{\left[j\theta(T_{1})\right]}, \label{eq30}
\end{eqnarray}
where $p$ and $\theta$ are real quantities and stand respectively for the amplitude and phase
of oscillations. After injecting (\ref{eq30}) into (\ref{eq28}),  we obtain
\begin{eqnarray} 
 -{p'} -j{\theta '}p+\frac{1}{2}p-
 \frac{(a+3b)p^3}{8}-\frac{E}{2}e^{j(\sigma T_{1}-\theta(T_{1}))}&=&0. \label{eq31}
\end{eqnarray}
We separate real and
imaginary terms and obtain the following coupled flow for the amplitude and phase:
 \begin{eqnarray}
\left\{
 \begin{array}{cl}
-{p'} +\frac{1}{2}p- \frac{(a+3b)p^3}{8}= \frac{E}{2} \cos {\Phi}   \label{eq33}         
\\
\\
-{\Phi '}p+ p\sigma = \frac{E}{2} \sin {\Phi},  
\end{array}
\right
.\end{eqnarray}
where the prime denotes the derivative with respect to $T_{1}$ and 
$\Phi = {\sigma} {T_{1}} - {\theta}(T_{1})$. For the steady-state
conditions $({p'} = {\Phi'} = 0 \Leftrightarrow p=p_{0}, \Phi=\Phi_{0} )$,  the following nonlinear algebraic equation is obtained:
\begin{eqnarray}
 \left(\frac{1+4\sigma^2} {4}\right){p_{0}}^2-\frac{a+3b}{8}{p_{0}}^4 +\frac{(a+3b)^2}{64}{p_{0}}^6=\frac{E^2}{4}, \label{eq34}
\end{eqnarray}
where ${p_{0}}$ and ${\Phi_{0}}$ are respectively the values of 
$p$ and $\Phi$ in the steady-state. 
 Eq.(\ref{eq34}) is the equation of primary resonance flow.
 Now,  we study the  stability of the precess,  we assume that each equilibrium state 
 is submitted to a small pertubation as follows
 \begin{eqnarray}
\left\{
 \begin{array}{cl}
p=p_{0}+p_{1}\label{eq35}         
\\
\\
\Phi=\Phi_{0}+\Phi_{1}\label{eq36}
\end{array}
\right
.\end{eqnarray}
where $p _{1}$ and $\Phi_{1}$ are slight variations. 
Inserting  the  equations (\ref{eq35}) into (\ref{eq33})   after some algebraic
and canceling nonlinear
terms enable us to obtain

 \begin{eqnarray}
{p'}_{1}&=&-\frac{1}{2}\left[1+ 3\frac{(a+3b)p_{0}^2}{4}\right]p_{1} -p_{0}\sigma \Phi_{1}, \label{eq37}         
\\
{\Phi '}_{1}&=& \frac{\sigma}{p_{0}}p_{1} -\frac{1}{2}\left[1- \frac{(a+3b)p_{0}^2}{4}\right] {\Phi}_{1}  \label{eq38}
\end{eqnarray}
The stability process depends on the sign of eigenvalues $\varGamma$ 
of the equations (\ref{eq37}) and (\ref{eq38}) which are given through
the following characteristic equation
\begin{eqnarray}
 \varGamma^{2}+2Q\varGamma+R&=&0, \label{eq48}
\end{eqnarray}
where
 \begin{eqnarray}
\left\{
 \begin{array}{cl}
Q=\frac{1}{2}\left(1+ \frac{(a+3b)p_{0}^2}{4}\right) \nonumber        
\\
\\
R=  \frac{1}{4}\left(1+ 3\frac{(a+3b)p_{0}^2}{4}\right)\left(1- \frac{(a+3b)p_{0}^2}{4}\right)-
\sigma^2.
\end{array}
\right
.\end{eqnarray}
Since $ Q > 0$,  the steady-state solutions are stable if $R > 0$ and unstable otherwise.
\begin{figure}[htbp]
\begin{center}
 \includegraphics[width=12cm, height=8cm]{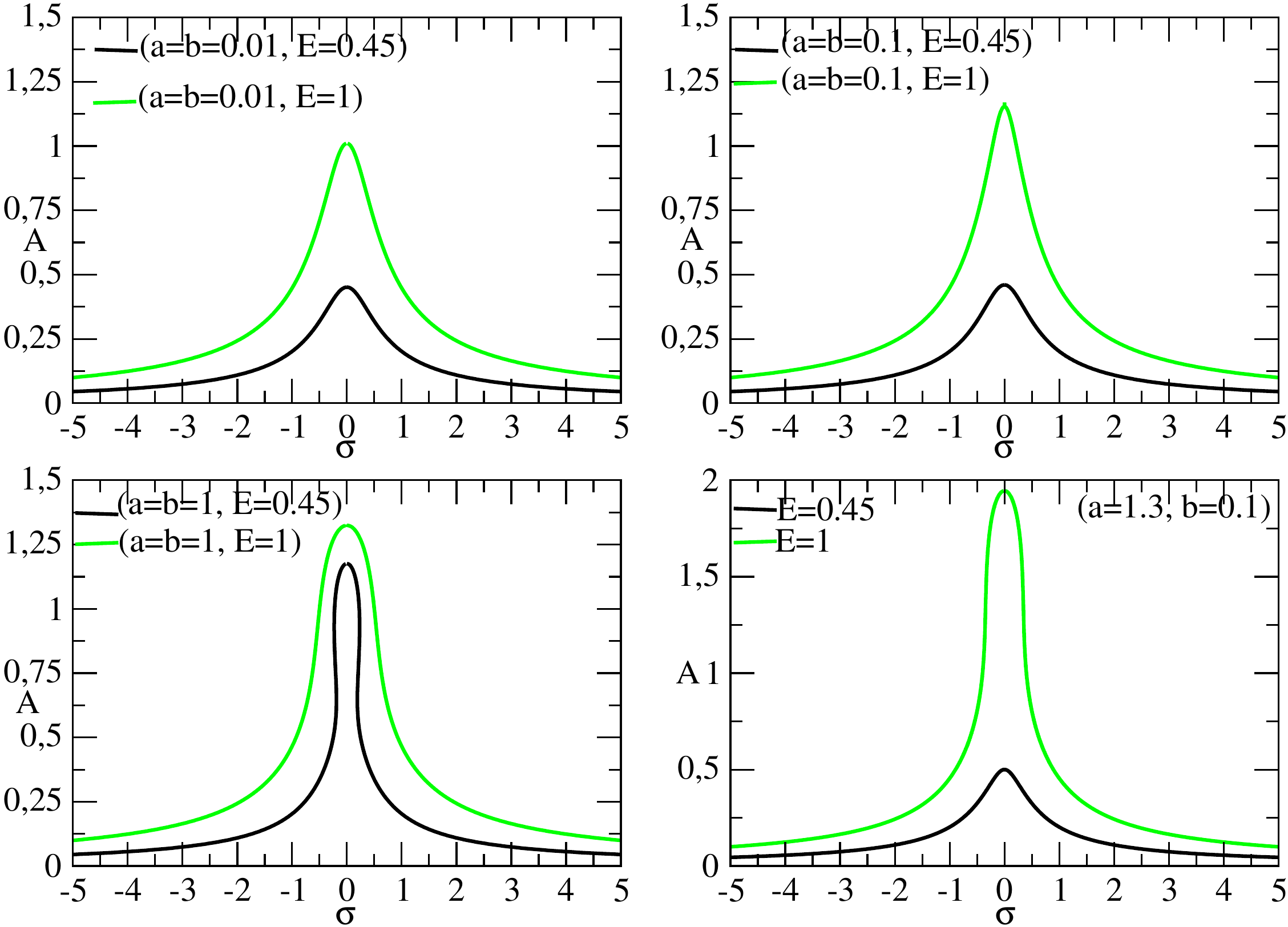}
\end{center}
\caption{Effects of the parameters $a, b, E $ on primary resonance curves}
\label{fig:3}
\end{figure}
\begin{figure}[htbp]
\begin{center}
 \includegraphics[width=12cm, height=5cm]{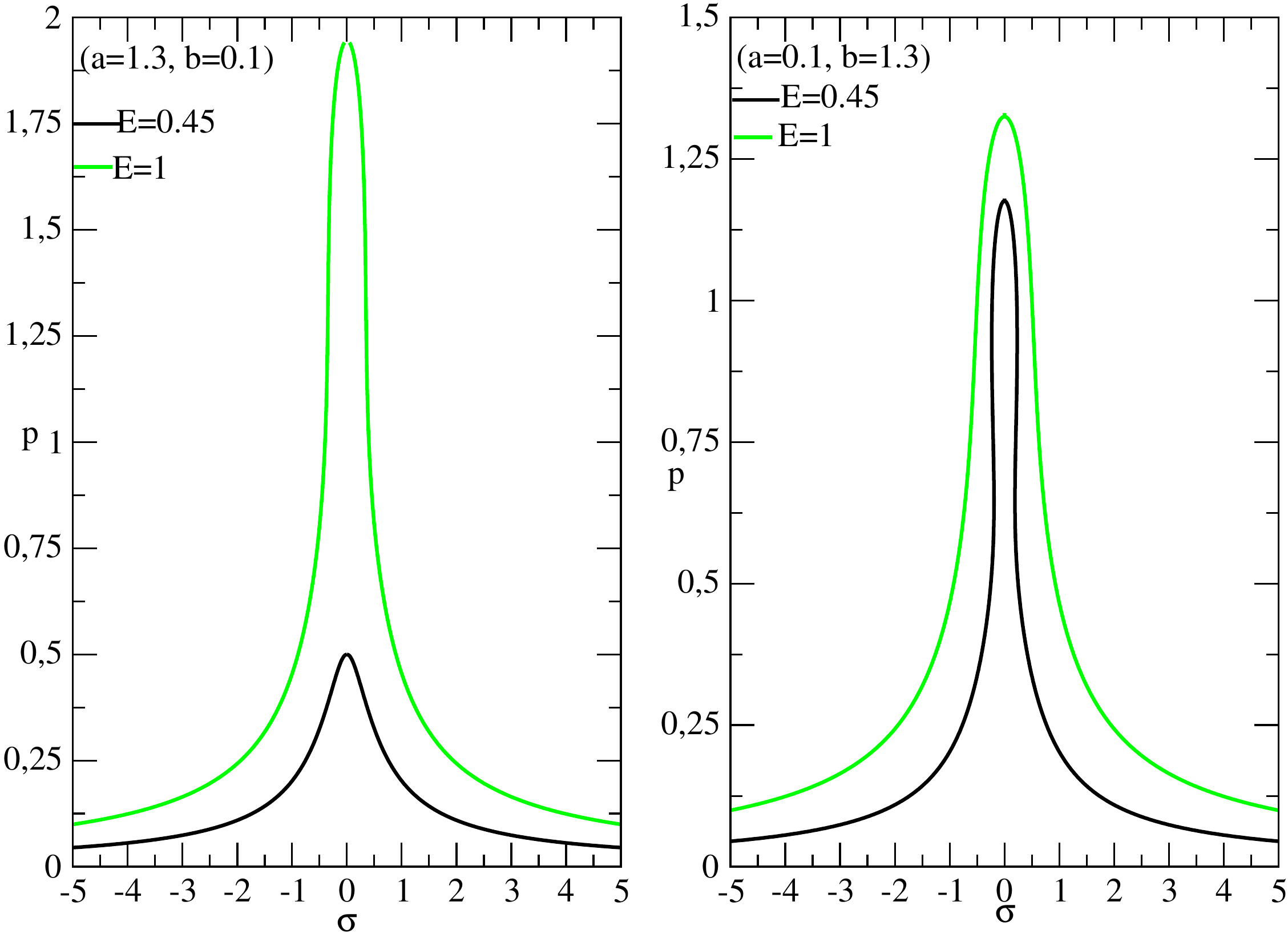}
\end{center}
\caption{Effects of  the parameters $a, b, E $ on the primary resonance curves}
\label{fig:4}
\end{figure}
\newpage
The figures \ref{fig:3} and \ref{fig:4}
display the primary resonance curves obtained from (\ref{eq34}) for 
different values of the parameters $a, b, E$. We obtained the linear resonances
curves and we found that resonance amplitude increase when the external forced amplitude increase. Through these figures we found also the peak value
of resonance amplitude is higher when $a+3b>1$ than $a+3b<1$. 
\begin{figure}[htbp]
\begin{center}
 \includegraphics[width=12cm,height=8cm]{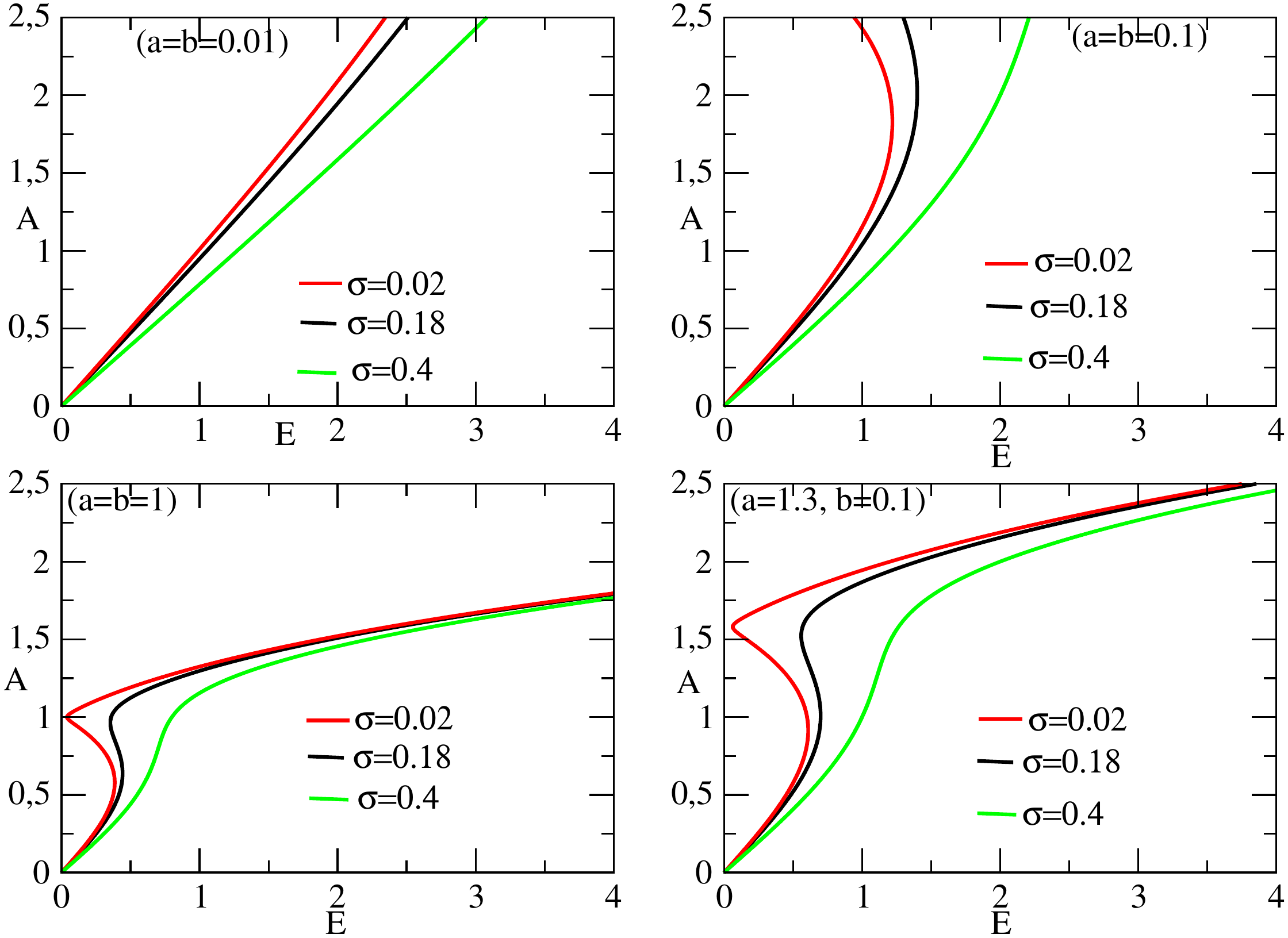}
\end{center}
\caption{Effects of  the parameters $a, b, \sigma $ on the amplitude-response curves}
\label{fig:5}
\end{figure}

The figure \ref{fig:5}
displays the amplitudes response curves obtained from (\ref{eq34}) for 
different values of the parameters $a, b, \sigma$. We obtained  for $a+3b\ll1$  
the amplitude-response  that is linear function of $E$
and the slope decreases as the $\sigma$ increases. The  hystereris's phenomena  
appears when $a+3b$ is important for small value 
of $\sigma$ and disappears when $\sigma$ increases. 

\subsection{Superharmonic and subharmonic oscillations}
When the amplitude of the sinusoidal external force is large,  other type of oscillations can be
displayed by the model,  namely the superharmonic and the subharmonic oscillatory states. It is
now assumed that $E \propto \varepsilon^{0}\tilde{E}$ and therefore,  one obtains the following equations at 
different order of $\varepsilon$.
In order ${\varepsilon}^{0}$, 
 \begin{equation}
  {D^2}_0 x_{0}+x_{0}=E\sin({\Omega T_0}).\label{eq49} 
  \end{equation}
In order $\varepsilon^1$, 
\begin{eqnarray}
 D^{2}_{0}x_{1}+x_{1}&=&-2D_{1}D_{0}x_ {0}+\left( 1-ax_{0}^2-b(D_{0}x_{0})^2\right)D_{0}x_{0}.\label{eq50}
\end{eqnarray}

The general solution of Eq.(\ref{eq50}) is
\begin{eqnarray}
\left\{
 \begin{array}{cl}
x_{0}=A(T_{1})e^{jT_{0}}+{\Lambda}e^{j{\Omega}T_{0}}, \label{eq54}       
\\
\\
\mbox{with} \qquad \Lambda=\frac{E}{2(1-{\Omega}^{2})}
.
\end{array}
\right
.\end{eqnarray}
Substituting the general solution $x_{0}$ into Eq. (\ref{eq50}),  after some
algebraic manipulations,  we obtain
\begin{eqnarray}
  D^{2}_{0}x_{1}+x_{1}&=&j\left[-2 A'+A\left(1 - \left(a+3b\right)|A|^2\right)-2A\Lambda^2\left(a+3b\Omega^2\right) \right] e^{jT_{0}}+\cr
  &&+j\left[{\Lambda}\Omega-2\left(a +3b\right)|A|^{2}\Lambda\Omega-\Lambda^3\Omega\left(a+3b\Omega^2\right)\right] e^{j\Omega T_{0}}+\cr  
  &&+j(b-a)A^3e^{3j T_{0}} +j \Lambda^{3} \Omega \left( b \Omega^2-a\right)e^{3j\Omega T_{0}}+ \cr
 &&+ j\left[-a\left(1+2 \Omega\right)+3b\Omega^2 \right]A \Lambda^2 e^{j(1+2\Omega)T_{0}}+\cr
  &&+ j\left[-a\left(2+\Omega \right) +3b\Omega \right]A^2\Lambda e^{j(2+\Omega)T_{0}}+\cr
  &&+j\left[a \left(2-\Omega \right)+3b\Omega \right] \bar{ A}^2\Lambda e^{j(\Omega-2)T_{0}}\cr
 &&+j\left[a\left(2\Omega-1\right)+ 3b\Omega \right]A{\Lambda}^{2} e^{j(1-2\Omega)T_{0}} +CC,\label{eq52}
\end{eqnarray}
where $CC$ represents the complex conjugate of the previous terms.

From Eq.(\ref{eq52}),  it comes that superharmonic and subharmonic states can be found from the
quadratic and cubic nonlinearities. The cases of superharmonic oscillation we consider is
$3\Omega = 1 + \epsilon \sigma$,  while the subharmonic 
oscillation to be treated is  $\Omega = 3 + \epsilon \sigma$.

\subsubsection{Superharmonic states}
For the first superharmonic states $3\Omega = 1 + \epsilon \sigma$,  equating resonant terms at $0$
 from Eq.(\ref{eq52}),  we obtain:
 
\begin{equation}
 -2A'+A\left[1-(a+3b)|A|^2-2\Lambda^2(a+3b\Omega^2)\right]+
 \Lambda^3 \Omega\left(\Omega^2 b -a\right)e^{j\varepsilon \sigma T_{0}}=0.\label{eq53}
\end{equation}
 Using (\ref{eq30}) and after some algebraic manipulations, 
we rewritte (\ref{eq53}) as follows
 \begin{eqnarray}
\left\{
 \begin{array}{cl}
p'=\frac{p}{2}\left[1- \frac{p^2}{4}(a+3b) -2\Lambda^2 (a+3b\Omega^2)\right]+
\Lambda^3 \Omega\left(b\Omega^2-a\right)\cos {\Phi}, 
\\
\\
p {\Phi '}=p\sigma -\Lambda^3 \Omega\left(b\Omega^2-a\right)\sin{\Phi},  \mbox{with} \qquad\Phi=\sigma T_{1}-\theta.
\end{array}
\right
.\end{eqnarray}
 The amplitude of oscillations of this superharmonic states 
 $({p'} = {\Phi'} = 0\Leftrightarrow p=p_{0}, \Phi=\Phi_{0} )$  is governed by
 the following nonlinear algebraic equation

 \begin{eqnarray}
\left\{
 \begin{array}{cl}
\frac{(a+3b)^2}{64}p_{0}^6-\mu \frac{a+3b}{4}p_{0}^4+\left(\mu^2+\sigma^2\right)p_{0}^2-
 \Lambda^6 \Omega^2\left( a-b\Omega^2\right)^2=0, \label{eq55}       
\\
\\
\mbox{with} \qquad \mu=\frac{1}{2}-\Lambda^2\left( a+3\Omega^2 b\right).
\end{array}
\right
.\end{eqnarray}
\newpage
\begin{figure}[htbp]
\begin{center}
 \includegraphics[width=12cm,  height=5cm]{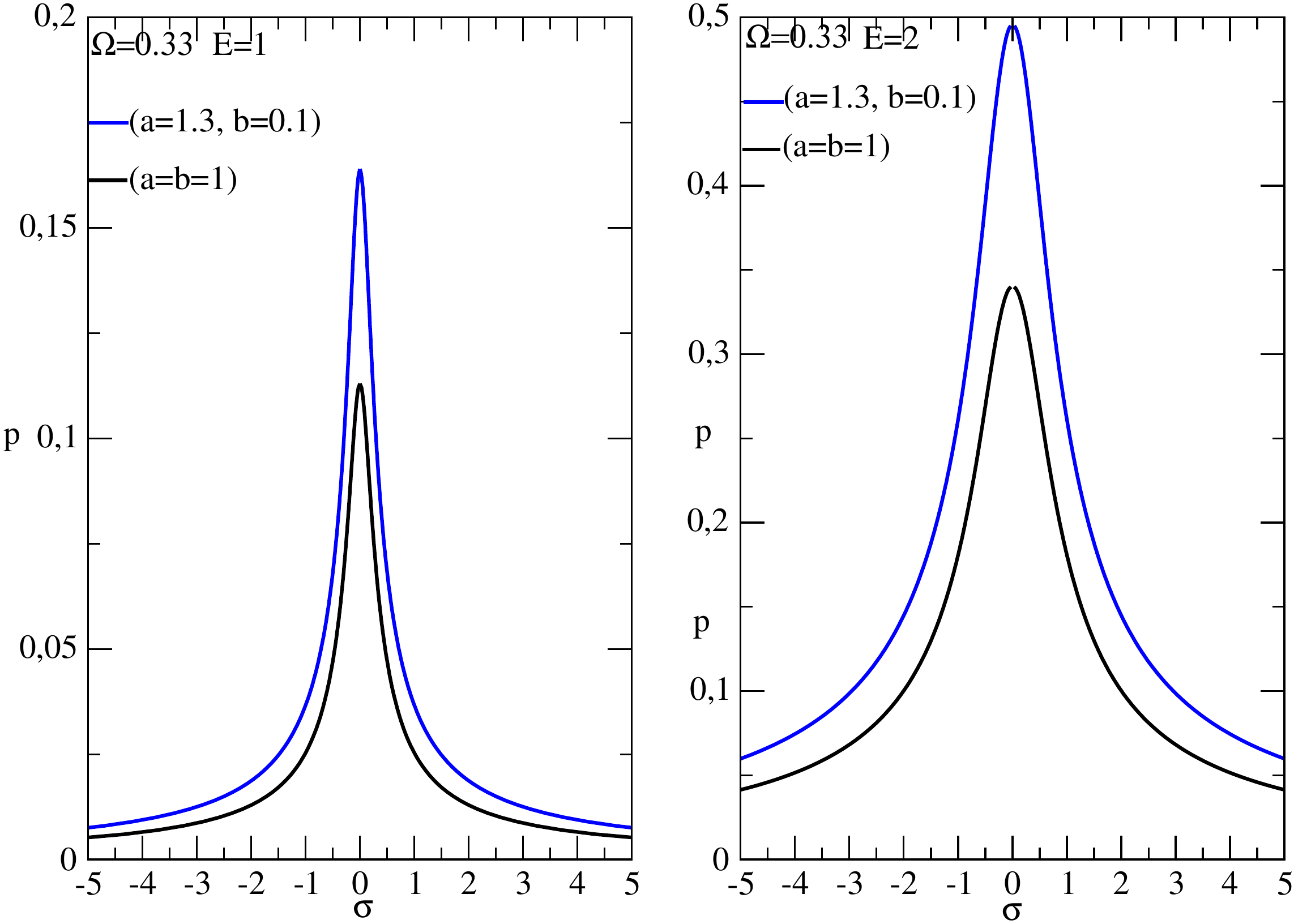}
\end{center}
\caption{Effects of  the parameters $a, b, E $ on the superharmonic resonance curves. 
}
\label{fig:6}
\end{figure}
From the figure (\ref{fig:6}) we note that in superharmonic states, the  resonance curve
is linear and the amplitude increases as the external force amplitude $E$ increases we note also, this 
amplitude deascreases as  $a+3b$  increases.  

\subsubsection{Subharmonic states}
For the first subhamonic states $\Omega = 3 + \epsilon \sigma$,  equating
resonant terms at $0$ from Eq.(\ref{eq52}),  we obtain:
\begin{eqnarray}
 2A'&=&
 A\left[1-(a+3b)|A|^2-2\Lambda^2(a+3b\Omega^2)\right]+\cr
&& \left[a \left(2-\Omega \right)+3b\Omega \right] \bar{ A}^2\Lambda e^{j\varepsilon \sigma T_{0}}.\label{eq56}
\end{eqnarray}
Using (\ref{eq30}) and after some algebraic manipulations, we rewritte (\ref{eq56}) as follows
\begin{eqnarray}
\left\{
 \begin{array}{cl}
p'=\frac{p}{2}\left[1- \frac{p^2}{4}(a+3b) -2\Lambda^2 (a+3b\Omega^2)\right]+
\left[a \left(2-\Omega \right)+3b\Omega \right]\frac{p^2 \Omega}{4}\cos {\varphi} \label{eq57}       
\\
\\
 \frac{\varphi'p}{3}-\frac{p\sigma}{3}=-\left[a \left(2-\Omega \right)+3b\Omega \right]\frac{p^2\Omega}{4}\sin {\varphi},
  \mbox{with} \qquad\varphi=\sigma T_{1}-3\theta.
\end{array}
\right
.\end{eqnarray}
The amplitude of oscillations of this subharmonic states 
 $({p'} = {\varphi'} = 0\Leftrightarrow p=p_{0}, \varphi=\varphi_{0} )$  is governed by
 the following nonlinear algebraic equation
 \begin{eqnarray}
\left\{
 \begin{array}{cl}
\frac{(a+3b)^2}{64}p_{0}^4 -\left(4\mu \left(a+3b\right)+  
\Lambda^2 \left[a \left(2-\Omega \right)+3b\Omega \right]^2\right)\frac{p_{0}^2}{16}+(\mu^2+\frac{\sigma^2}{9})=0, \label{eq58}       
\\
\\
\mbox{with}\qquad \mu=\frac{1}{2}-\Lambda^2\left( a+3\Omega^2 b\right).
\end{array}
\right
.\end{eqnarray}

\begin{figure}[htbp]
\begin{center}
 \includegraphics[width=12cm,  height=6cm]{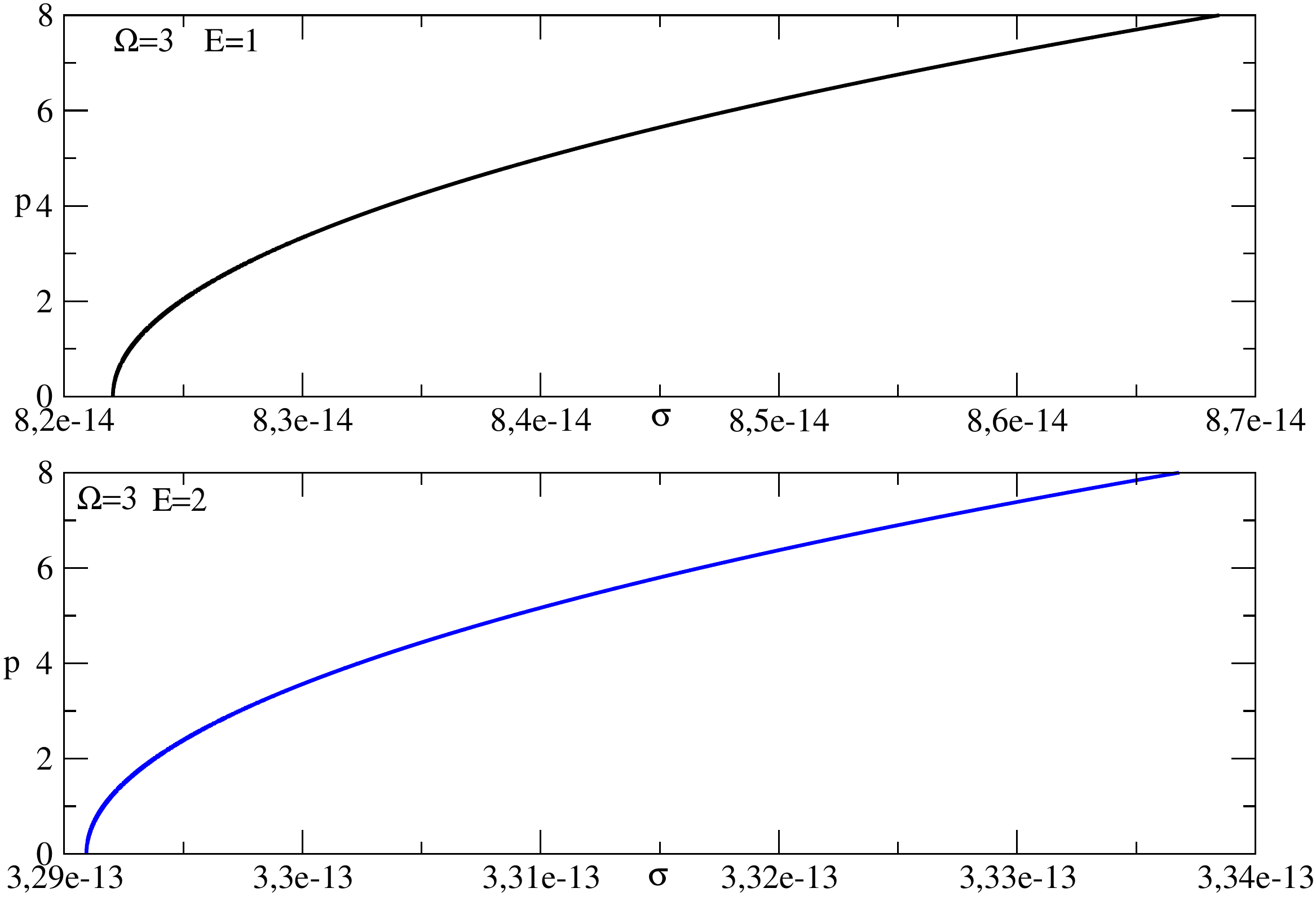}
\end{center}
\caption{Effects of  the parameters $E$ on subharmonic resonance  curves with $a=b=1$.}
\label{fig:7}
\end{figure}

 From the Fig. (\ref{fig:7}) we note a high sensibility of amplitude to the frequencie. 
\newpage
 \section{Bifurcation and Chaotic behavior}
The aim of this section is to find some bifurcation structures in the nonlinear dynamics of forced Van der Pol generalized oscillator 
described by equation (\ref{eq1}) for resonant states since they are in interest for the system. For this purpose,  
we numerically solve this equation using the fourth-order Runge Kutta algorithm \cite{15} and plot the resulting bifurcation
 diagrams and the variation of the corresponding largest Lyapunov exponent as the amplitude $E$,  the parameters of nonlinearity 
$ a,  b$  varied. The stroboscopic time period used to map various transitions
which appear in the model is $ T=\frac{2\pi}{\Omega}$. 

\begin{figure}[htbp]
\begin{center}
\includegraphics[width=12cm,  height=8cm]{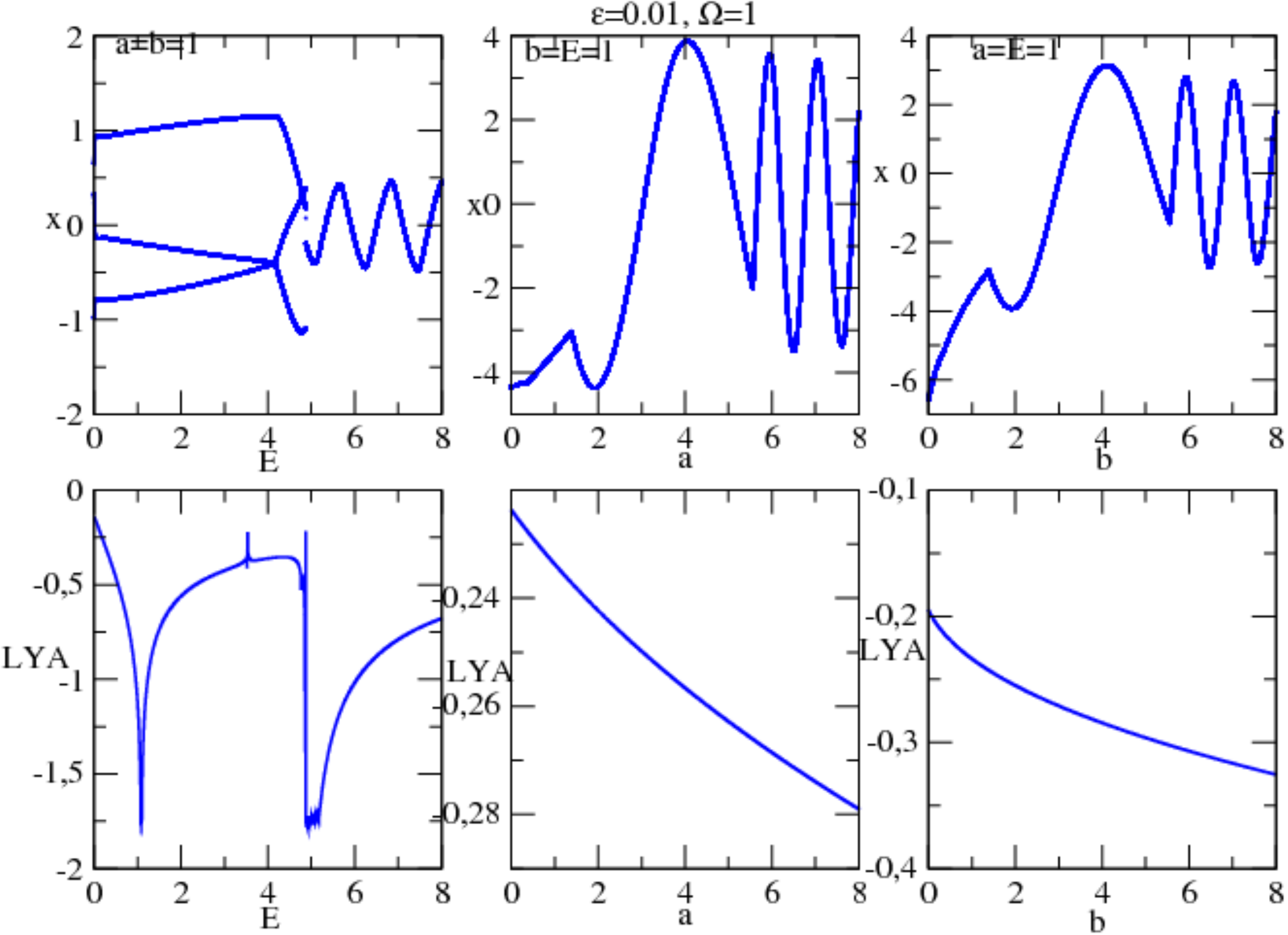}
\end{center}
\caption{Bifurcation  diagram (upper frame) and Lyapunov exponent (lower frame) 
for $\varepsilon=0.01, \Omega=1$ primary resonance case.}
\label{fig:8}
\end{figure}

\begin{figure}[htbp]
\begin{center}
 \includegraphics[width=12cm,  height=8cm]{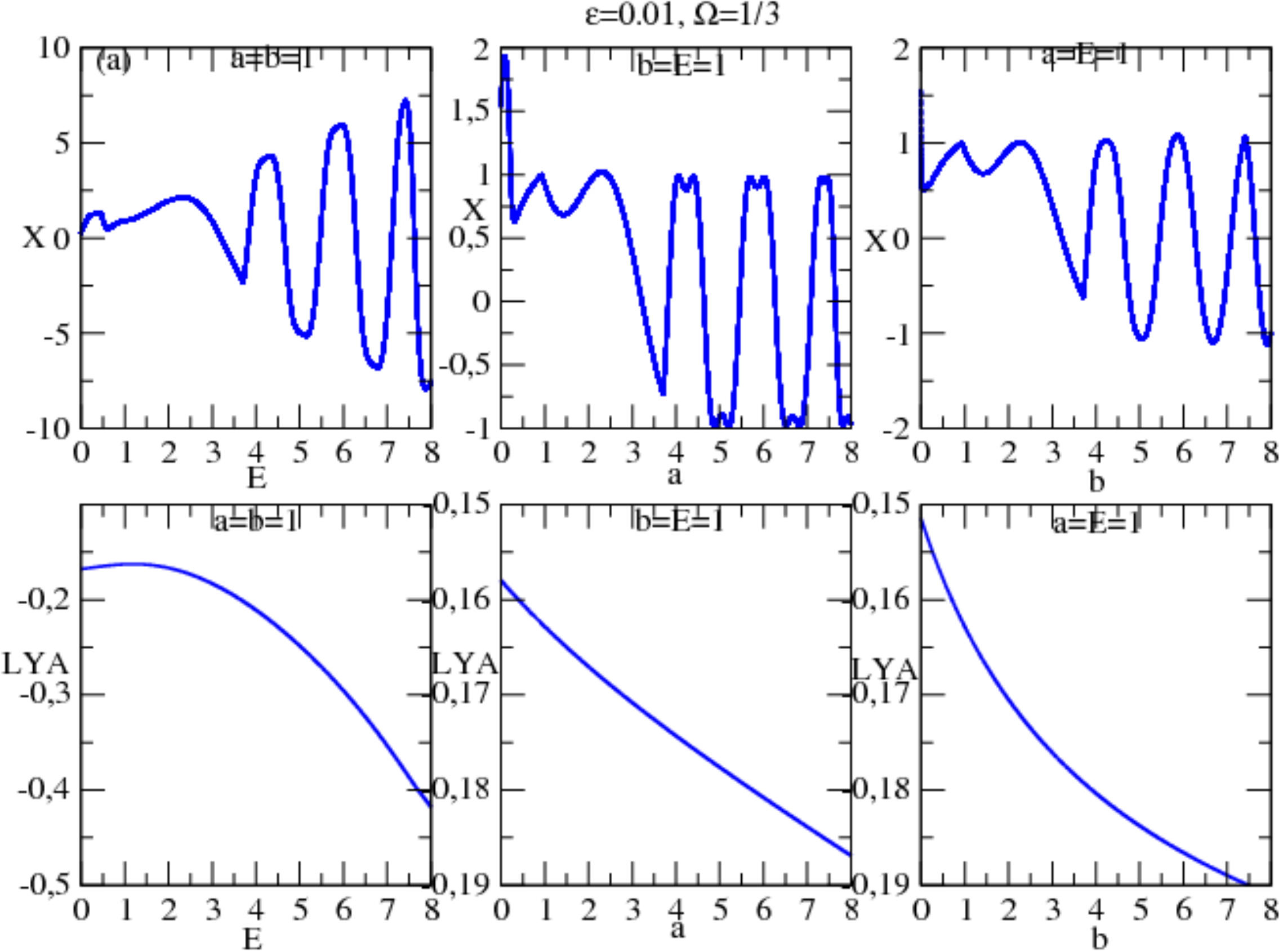}
\end{center}
\caption{Bifurcation  diagram (upper frame) and Lyapunov exponent (lower frame) 
for $\varepsilon=0.01, \Omega=\frac{1}{3}$ supharmonic case.}
\label{fig:9}
\end{figure}

\begin{figure}[htbp]
\begin{center}
 \includegraphics[width=12cm,  height=8cm]{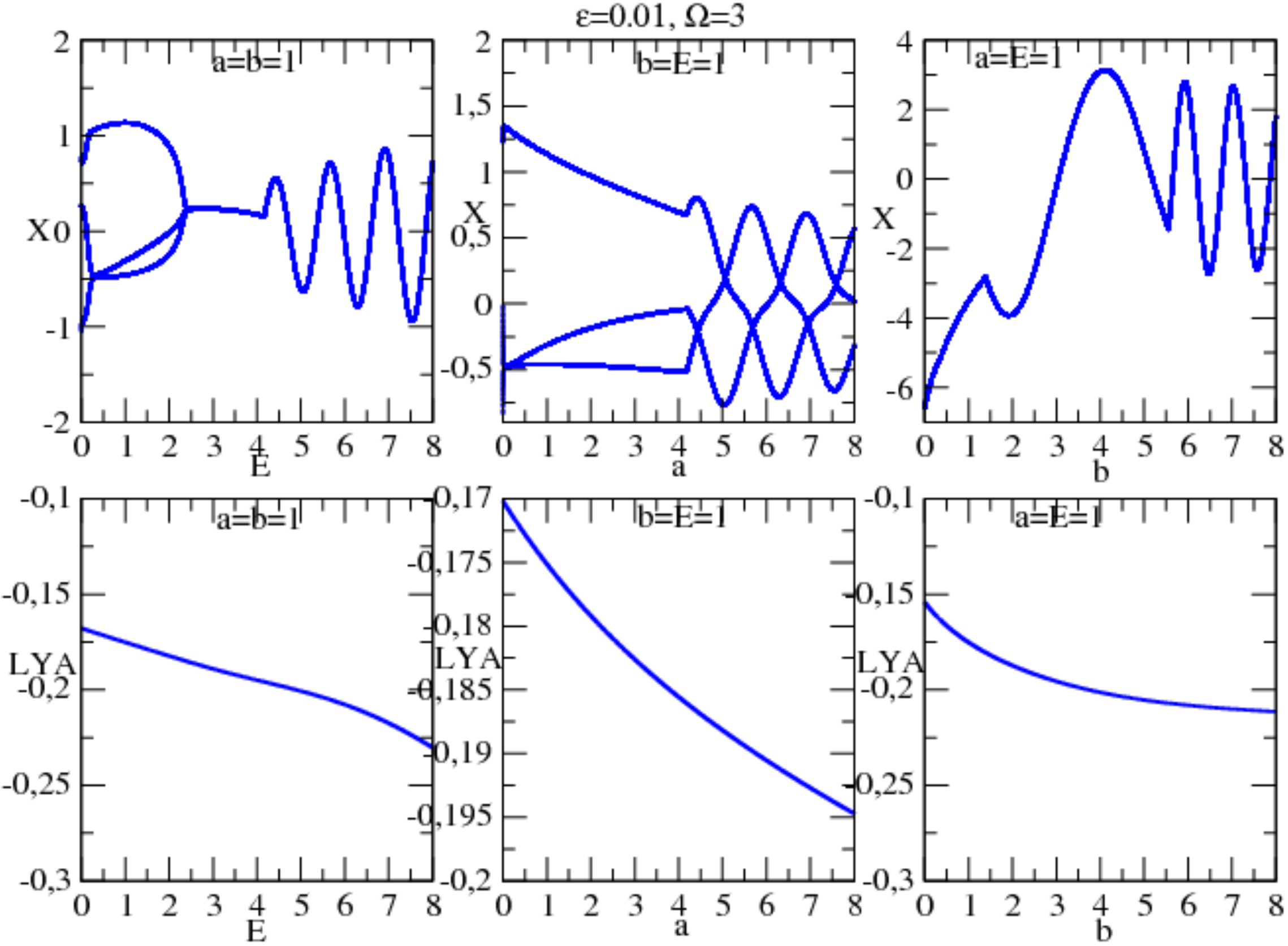}
\end{center}
\caption{Bifurcation  diagram (upper frame) and Lyapunov exponent (lower frame) 
for $\varepsilon=0.01, \Omega=3$ subharmonic case.}
\label{fig:10}
\end{figure}

\begin{figure}[htbp]
\begin{center}
 \includegraphics[width=12cm,  height=8cm]{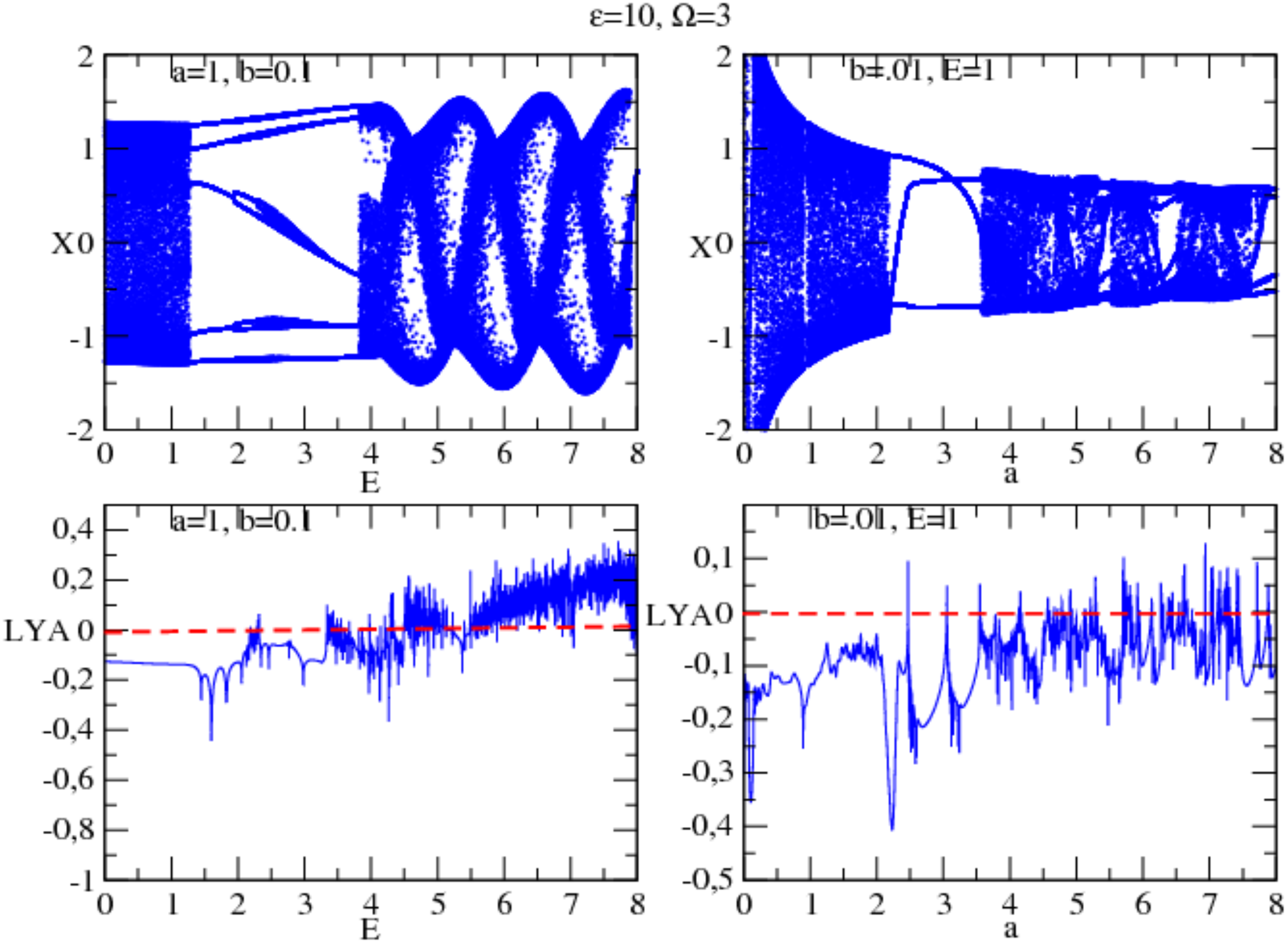}
\end{center}
\caption{Bifurcation  diagram (upper frame) and Lyapunov exponent (lower frame) 
for $\varepsilon=10, \Omega=3$  Subharmonic states case.}
\label{fig:11} 
\end{figure}

\begin{figure}[htbp]
\begin{center}
 \includegraphics[width=12cm,  height=8cm]{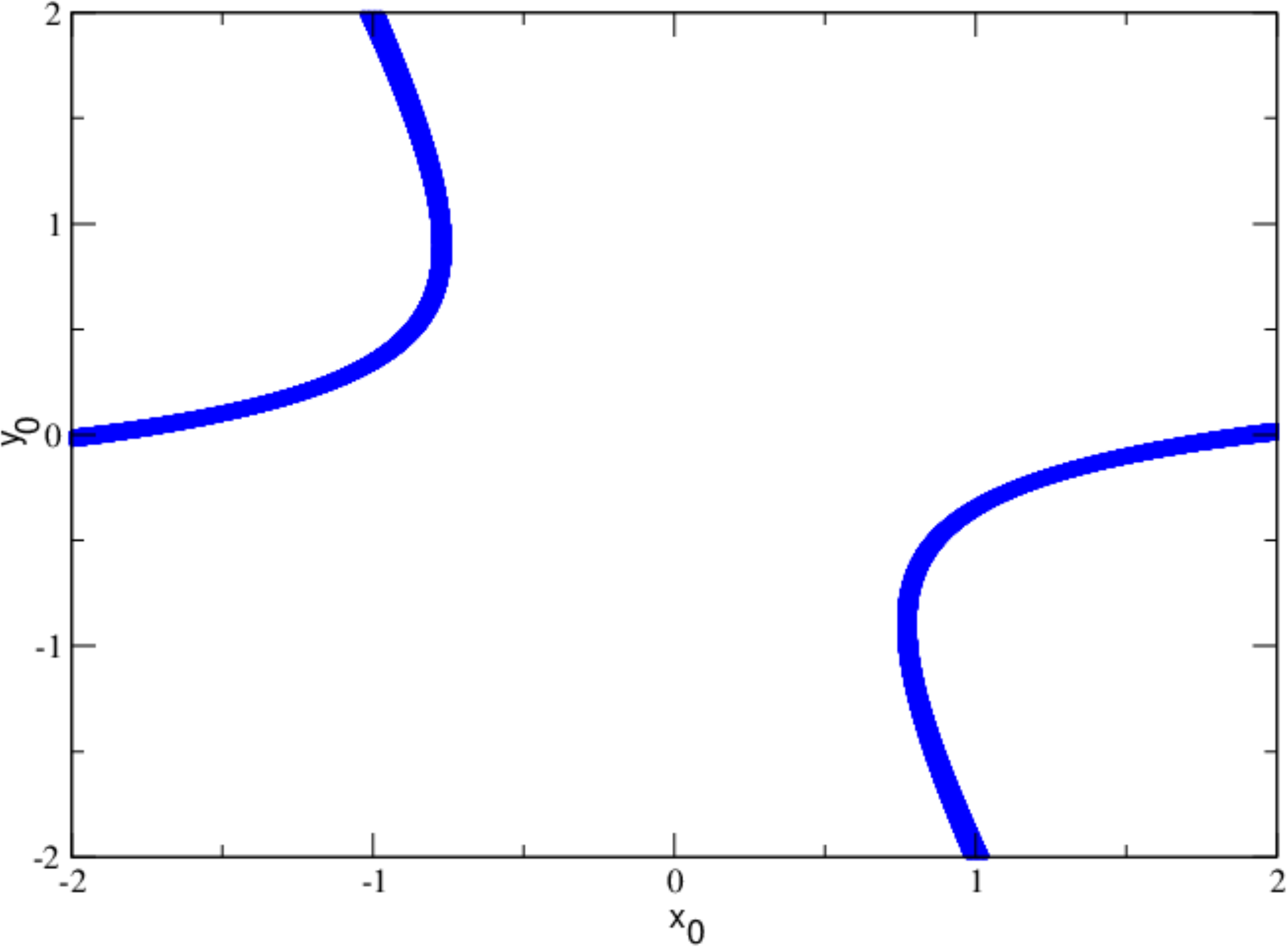}
\end{center}
\caption{Chaoticity basin  for  parameters values in  figure \ref{fig:9} with $E=1$}
\label{fig:16} 
\end{figure}

\begin{figure}[htbp]
\begin{center}
 \includegraphics[width=12cm,  height=8cm]{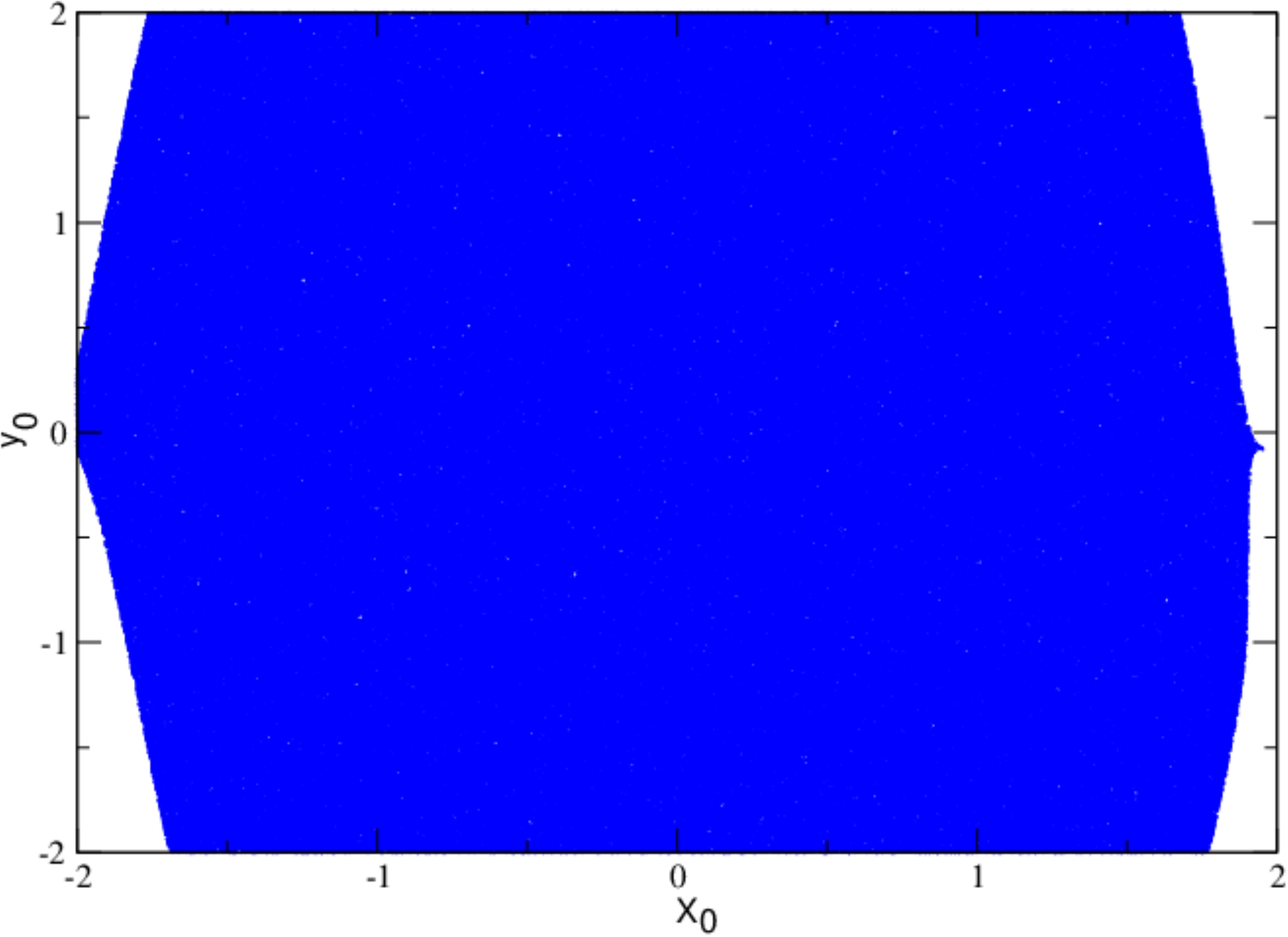}
\end{center}
\caption{Chaoticity basin  for  parameters values in  figure \ref{fig:11} with $E=7.5$}
\label{fig:17} 
\end{figure}

The largest Lyapunov exponent which is used here as the instrument to measure the rate of chaos in the system is defined as
\begin{equation}
 Lya=\lim_{t \rightarrow \infty}\frac{ln \sqrt{dx^2+d\dot x^2}}{t}\label{eq70}
\end{equation}
where $dx$ and $d\dot x$ are respectively the variations of $x$ and $\dot x$.
Initial condition that we used in the simulations of this section is $(x_0, \dot x_0)=(1, 1)$.
For the set of parameters $a =\{1, 6\}$,  $b=\{1, 6\}$, $E=\{1, 6\}$,  $\epsilon = \{0.01, 10\}$,  $\Omega =\{1,\frac{1}{3}, 3\}$,   
 the  bifurcation and Lyapunov exponents diagrams   for  primary, superharmonic and subharmonic resonances 
 are plotted  respectively in Figs. (\ref{fig:8}), (\ref{fig:9}) and ( (\ref{fig:10}),  (\ref{fig:11})).
 The bifurcation  diagrams  are in upper frame and its corresponding Lyapunov exponents  are in lower frame.  
 From the  diagrams it is found that the model can switch from periodic to 
 quasi-periodic, nonperiodic and chaotic oscillations. Since the model is highly sensitive to the initial conditions, it can leave a quasi-periodic
state for a chaotic state without changing the physical parameters. Therefore, its basin of attraction  has been plotted
(see Figs.(\ref{fig:16}), (\ref{fig:17}))
in order to situate some regions of the initial conditions for which chaotic oscillations are observed. 
From these figures, we conclude that chaos is more abundant in
the subharmonic resonant states than in the superharmonic and primary resonances. This confirms
what has been obtained through their bifurcation diagrams and Lyapunov exponent.

We noticed that for the small value of $\varepsilon$ the system is not chaotic.

\begin{figure}[htbp]
\begin{center}
 \includegraphics[width=12cm,  height=8cm]{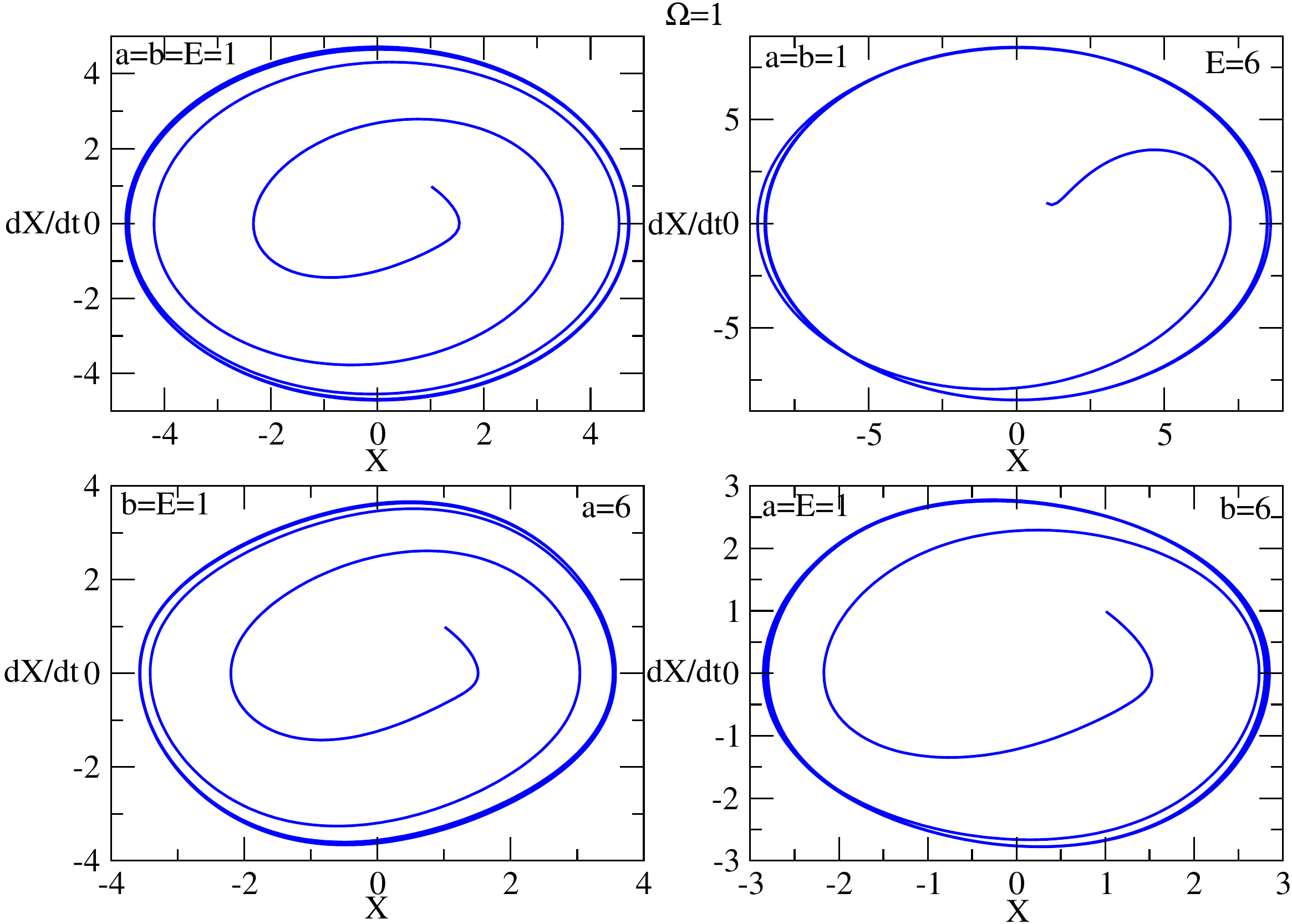}
\end{center}
\caption{ Phases diagram for  parameters values in  figure, $\varepsilon=0.01$, primary resonance case.}
\label{fig:12}
\end{figure}

\begin{figure}[htbp]
\begin{center}
 \includegraphics[width=12cm,  height=8cm]{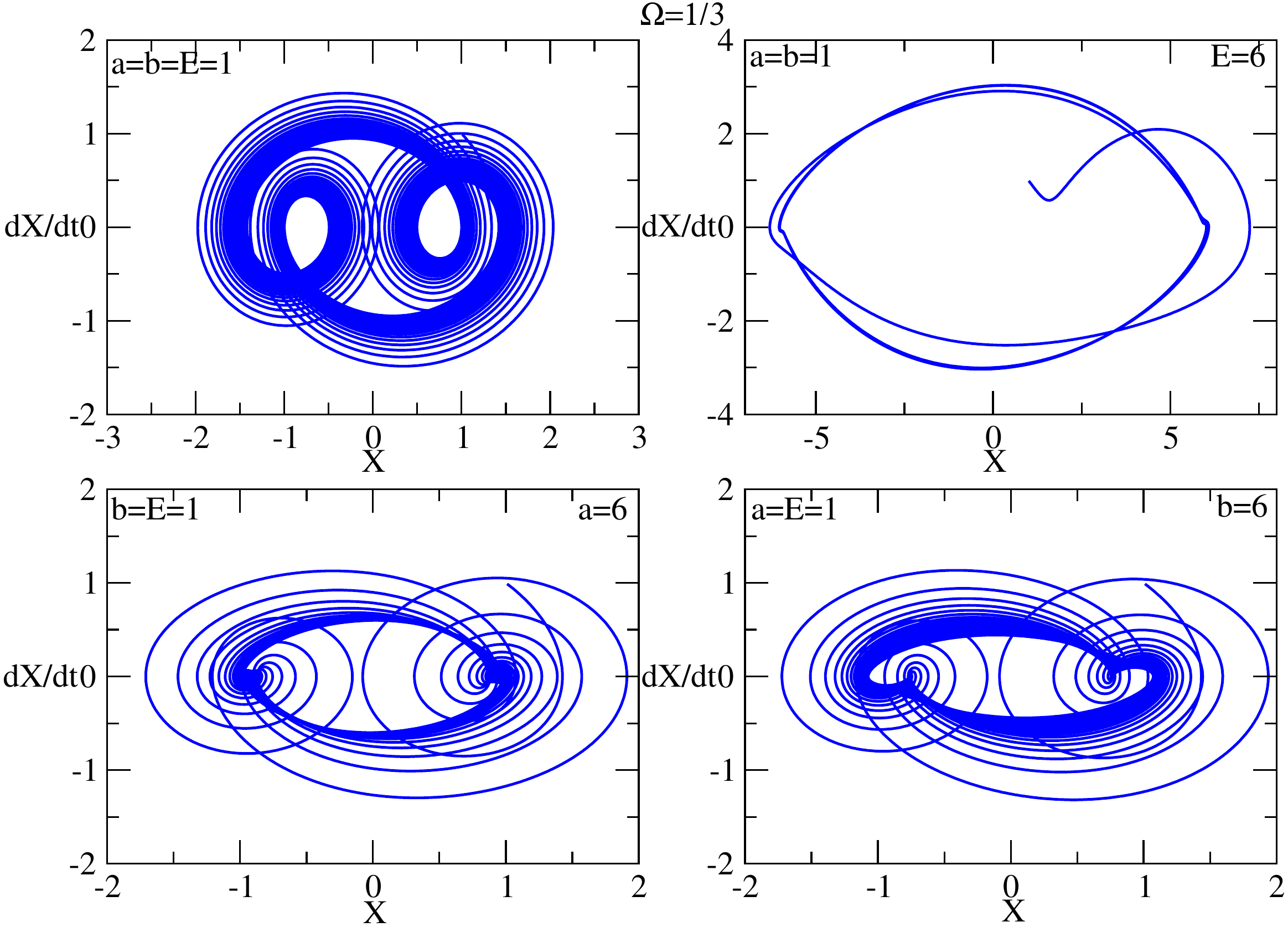}
\end{center}
\caption{Phases diagram for  parameters values in  figure, $\varepsilon=0.01$: Superharmonic states case.}
\label{fig:13}
\end{figure}

\begin{figure}[htbp]
\begin{center}
 \includegraphics[width=12cm,  height=8cm]{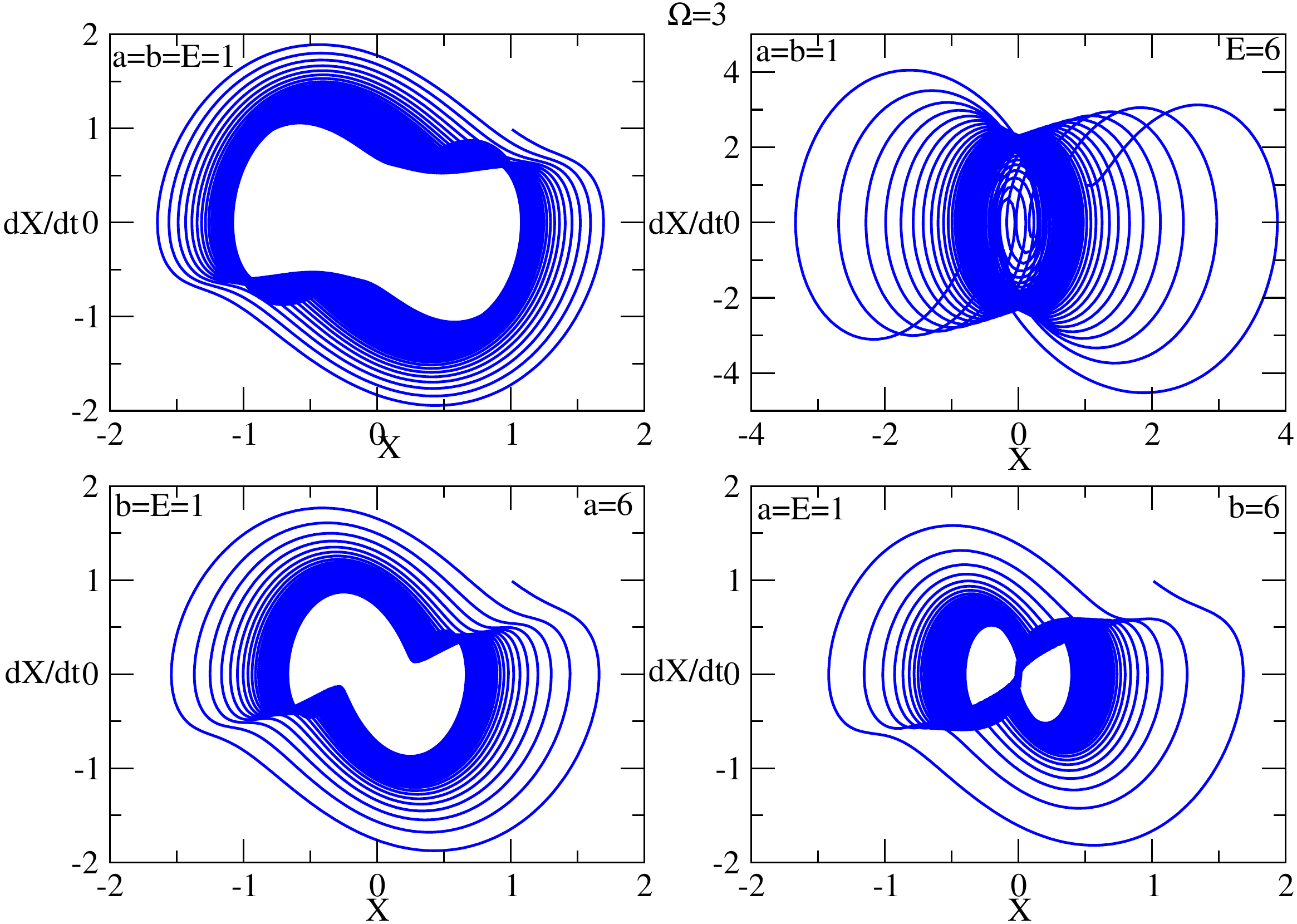}
\end{center}
\caption{ Phases diagram for  parameters values in  figure, $\varepsilon=0.01$:  Subharmonic states case.}
\label{fig:14} 
\end{figure}

\begin{figure}[htbp]
\begin{center}
 \includegraphics[width=12cm,  height=8cm]{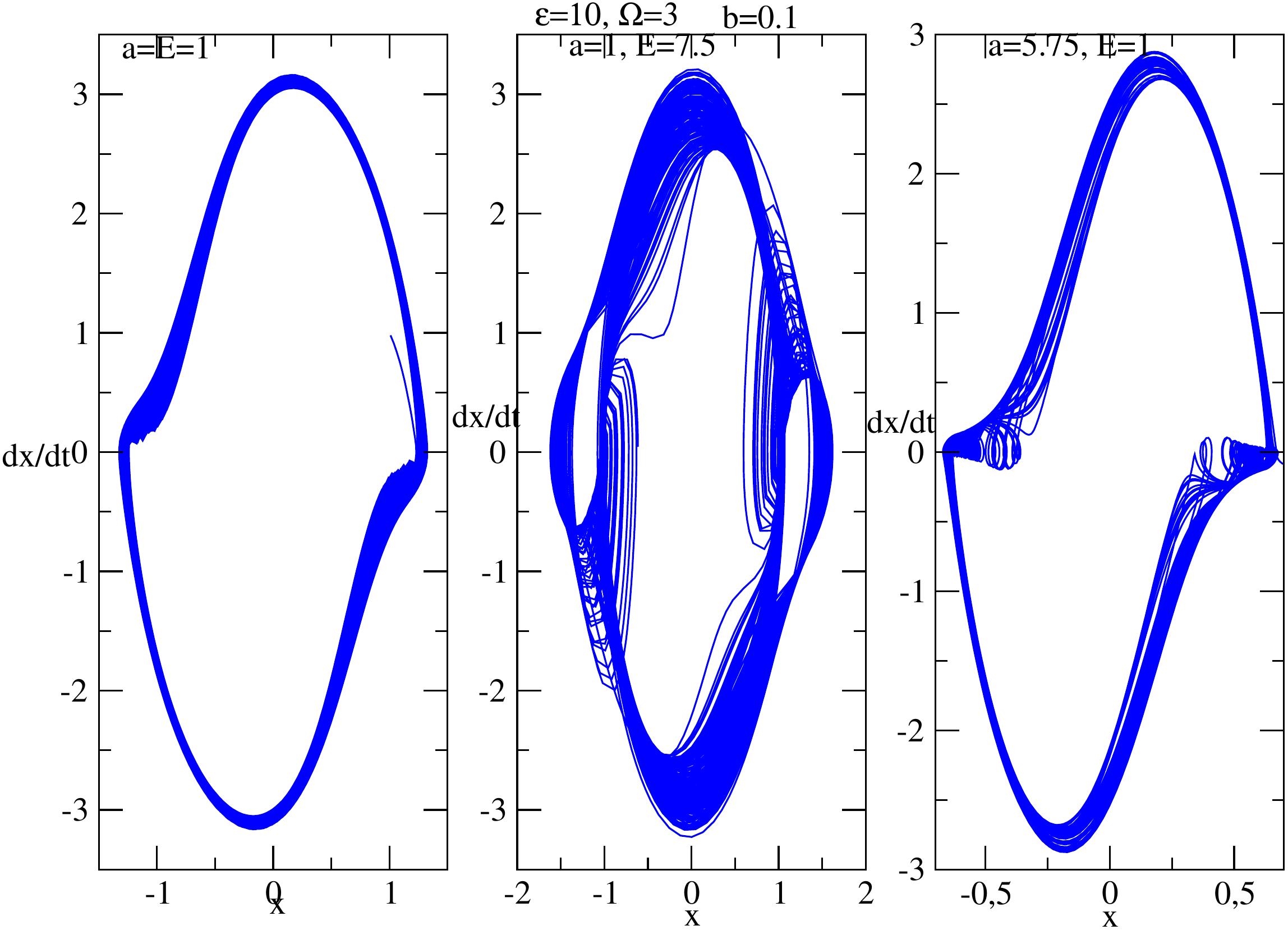}
\end{center}
\caption{Phases diagram for  parameters values in  figure, $\varepsilon=10, \Omega=3$.}
\label{fig:15} 
\end{figure}

 In order to illustrate such situations,  we have represented the various phase portraits
 using the parameters of the bifurcation diagram for 
 which periodic, quasi-periodic and nonperiodic oscillations motions are observed in Figs.
(\ref{fig:12}), (\ref{fig:13}), (\ref{fig:14})and chaotic motions are observed in Fig.(\ref{fig:15})  for the  two  values of 
the parameters $a, b, E$ above. 
 From the  phases diagram we observed  appearance of the closed curve and
 the torus that confirm the two precedent types of motions 
predicted and the linear types of resonances seen in the third section.

\newpage
\section{Conclusion}

In this work,  we have studied the nonlinear dynamics for system oscillations
modeled by forced  Van der Pol generalized oscillator. In the harmonic case,  
the balance method has enabled us to derive the amplitude of harmonic oscillations,and the effects of the differents
parameters on the behaviors of model have been analyzed. 
For the resonant states case,  the response amplitude,  stability (for primary resonance case) have been derived by
using multiple time-scales method and pertubation method.
It appears the  first-orders superharmonic and subharmonic resonances.
The effects of differents parameters on these resonances are been found and we noticed that the Van der Pol, Rayleigh 
 parameters and the external force amplitude have
several action on the  amplitude reponse  and the resonances curves. 
The influences of these parameters on the resonant,  hysteresis and jump phenomena have been highlighted. 
Our analytical results have been confirmed by numerical simulation. Various bifurcation structures showing different 
types of transitions from quasi-periodic motions to periodic and  the beginning of chaotic motions
have been drawn and the influences of different parameters on these motions have been study. It is noticed 
that behaviors of system have been controlled by the  parameters $ a, b,  E$ and  but also 
the damping parameter $\epsilon$.  We conclude that chaos is more abundant in
the subharmonic resonant states than in the superharmonic and primary resonances. This confirms
what has been obtained through their bifurcation diagrams and Lyapunov exponent.
  The results show a way to predict admissible values of the signal amplitude for a corresponding set of parameters. 
This could be helpful for the experimentalists who are interested in trying to stabilize such a system with external forcing.
For practical interests,
it is useful to develop tools and to find ways to control or suppress such undesirable regions. This will be also  useful to 
control high amplitude of oscillations obtained 
and which are generally source of instability in physics system.

\section*{Acknowlegments}
The authors thank IMSP-UAC and Benin gorvernment for financial support.

\end{document}